\documentstyle[preprint,aps,epsfig]{revtex}
\newcommand{\ba}{\begin{eqnarray}}
\newcommand{\ea}{\end{eqnarray}}
\newcommand \Pomeron {I\!\!P}
\begin{document}
\title{Diffractive production of charm
in DIS
and  gluon nuclear shadowing}

\author{L.~Alvero}
\address{Department of Physics, Pennsylvania State University,
         University Park, PA 16802}
\author{L.~L.~Frankfurt}
\address{School of Physics and Astronomy, Tel Aviv University,
         69978 Tel Aviv, Israel}

\author{M.~I.~Strikman}
\address{Department of Physics, Pennsylvania
State University, University Park, PA 16802, USA,\\
and Deutsches Elektronen Synchrotron DESY, Germany\thanks{On leave of
absence from PSU.}}

\maketitle

\begin{abstract}
We evaluate nuclear shadowing of the total cross section of charm
particles production in DIS within the framework of
Gribov theory of nuclear shadowing
generalized to account for the QCD evolution. We use
as an input the recent QCD Pomeron parton density analysis of the
HERA diffractive data. Assuming that
the QCD factorization theorem is applicable to the
charm production off nuclei we also calculate shadowing of
the  gluon densities in nuclei
and find it sufficiently large for heavy nuclei:
$G_{A\sim 200}(x,Q^2)/AG_N(x,Q^2)
\sim 0.45-0.5 \cdot (A/200)^{-0.15}$
for $x \sim 10^{-3\div -4},  Q^2 \sim 20 \div 40 GeV^2$
to influence significantly the physics of heavy
ion collisions at LHC.
We evaluate also suppression of minijet and hidden charm production in
the central AA collisions.
We also discuss some
properties of the final states for $\gamma^*A $  processes
dominated by the scattering off
small  $x$ gluons like the high $p_t$ jet
and  charm production.
\end{abstract}
\newpage

\section{Introduction}
The aim of this paper is to
evaluate
gluon shadowing in nuclei in  the HERA
kinematics using generalised  Gribov theory of nuclear shadowing
and information on diffraction processes in DIS obtained at HERA.

The paper is organized as follows. In section 2 we review the Gribov 
theory of nuclear shadowing for $F_{2A}(x,Q^2)$ \cite{Gribov}
which established a connection of  this phenomenon with the process of 
the 
diffraction in the elementary process of the scattering of a 
projectile off a single nucleon (eq.(4)).
In essence, this connection is due to a possibility of very 
low momentum transfer to a nucleon in the diffraction processes comparable to 
internal momenta of nucleons in nuclei which leads to a significant
 interference 
of the amplitudes   of diffractive scattering off different nucleons of 
the nucleus.  We explain also that the Gribov theory takes into account 
only contribution of large longitudinal distances 
in deep inelastic scattering. Hence one has to take into account an 
additional QCD effect of the change of the average longitudinal
distances with increase of $Q^2$ at fixed $x$.
 This ammounts to taking 
into account  the leading twist QCD evolution of the parton densities
at $x \ge 0.02$ at $Q^2 \ge Q_0^2$. Numericaly this effect is very 
small in the kinematics which we study in this paper.

In section 3 we focus on shadowing in  the charm production 
off nuclei and gluon nuclear 
 shadowing.
We are able to  extend the Gribov theory to these processes due to
the recent 
systematic studies at HERA of the diffractive phenomena
in DIS, and in the real photon induced hard processes, see \cite{HERAdiff}.
One important observation beared by all analyses of the HERA data as
 well as of the hard hadronic diffraction data is the
dominance of the gluon degrees of freedom 
as compared to quark degrees of freedom (by a factor $\sim 5-6$).
This dominance at $Q^2\le 20 GeV^2$ is much 
stronger than in the case of small $x$ parton densities at $x \le
10^{-4}$
(a factor $\sim 2$).
Hence we come to {\it a qualitative conclusion that 
shadowing in the gluon channel should be 
larger ($\sim 2-3$) that in the charge parton channel}.


To quantify this conclusion we use the global
analysis of the diffractive HERA data \cite{ACW} wich 
has demonstrated that the scaling violation in $\sigma_{diff}(x,Q^2)$,
diffractive production of jets in the $\gamma p$ scattering, diffractive
production of charm in DIS can be described  based on the
leading twist factorization approximation  for the  parton densities of
diffraction which are often referred to as the Pomeron structure
functions. In this paper we focus on the $Q^2$ range
covered by the HERA diffractive experiments sensitive
to the gluon intiated diffraction: $Q^2 \sim 20 -50 $ GeV$^2$
and the $x \le 10^{-3}$ range where average longitudinal distances
in the $\gamma^* -A$ interactions are much larger than the nucleus
size. We will discuss the $Q^2$ dependence of shadowing for lower
 $Q^2$ and the
whole range of $x$ in
the subsequent paper \cite{FS982}.
Our results for the gluon shadowing are given by eqs.6, 9, 10.
they provide a direct connection between the gluon shadowing due to
the interaction with two nucleons and hard diffraction processes
with nucleons  domianted by the gluon degrees of freedom.
The Gribov theory does not produce unique predictions for the
contribution of the interactions where three or more nucleons are involved.
As a starting point we treat these interactions neglecting
effects of fluctuations of the strength of interaction of the projectile 
with a target. This corresponds to the quasieikonal approximation in
 the hadron-nucleus scattering.
The results of numerical studies in this approximation presented in Fig.2
confirm that shadowing in the gluon channel is indeed larger
(by about a factor of two for $Q^2 \sim 20 GeV^2$).
than the quark  shadowing.  

Next we analyse effects of the fluctuation of the strength of
 interaction. 
  Already simple dimensional arguments 
indicate that these fluctuations in our case should be large. 
We analyse this effect and find that indeed it 
becomes important in the case of suffiently large $A$ which ammounts to 
a violation of the quasieikonal
approximation and filtering out of the contribution
of the strongly interacting configurations.
However we find that even for central impact parameters for 
heavy nuclear targets ($A\le 250$)
 this effect remains a correction. One needs $A \ge 1000$
to reach the region where these effects dominate. 
In the end of   the paper we analyse the final states in the charm production.
We predict that for the case of scattering of heavy nuclei
coherent diffractive photoproduction 
of charm would constitute nearly half of the
total charm production cross section. We also predict strong
screening effects for coherent exclusive production of
vector mesons in DIS $Q^2 \le 20 GeV^2$ and $x \le 10^{-3}$.

\section{Shadowing of quark distributions within a nucleus.}

First let us summarize necessary elements of the Gribov theory
in the case of DIS scattering of a deuteron where calculation can be
performed in a model independent way.
Theoretical description of
nuclear shadowing is simplified in the nucleus rest frame. The starting
point is the presence of large coherence length in the interaction
of a $\gamma^*$ with a nucleon with mass $m_N$: a virtual photon
 with the "mass$^2$" $\sim -Q^2$
fluctuates into a  quark-gluon configuration of characteristic invariant
mass $M_X^2$
over the average longitudinal distances (in the nucleus rest frame)
see \cite{ioffe} and for further development see \cite{FS89}
and references wherein
\begin{equation}
l_{coh}= {1 \over  m_Nx(1+M_X^2/Q^2)} \ .
\label{lcoh}
\end{equation}
$l_{coh}$
corresponds to  the average distance over which
the  light-cone wave function of a
photon which includes radiation of partons is build..
In preQCD period it has been assumed that $M_{X}^2\approx Q^{2}$. In
this case the expression for coherence length obtaines famous form
\begin{equation}
l_{coh}= {1 \over 2 m_Nx} \ ,
\label{lcoh1}
\end{equation}
which by far exceed the nuclear radius even for heavy nuclei.
$l_{coh}\gg R_A$.

For this limit Gribov has established the connection between
the phenomenon of diffraction in the  scattering
off a nucleon and the presence of the nuclear shadowing
\cite{Gribov}. The key elements of the derivation are the
dominance of the vacuum exchange in the $t$-channel
of the amplitude of the $aN$ scattering, which is
well established now experimentally,  and the possibility to
describe  the deuteron and heavier  nuclei as  loosely bound systems
build of nucleons.
The difference in the scales characterizing a nuclear bound state
(average internucleon distances in nuclei $r_{NN}\approx 1.7 Fm$
and somewhat larger within the deuteron) and
the nucleon radius,  $r_{N}\approx 0.8 Fm$,
justifies taking into account only the  contribution of the nearest
pole in the nuclear vertex.
A similar
assumption is routinely made in the Glauber theory of the
intermediate energy hadron-nucleus
scattering which  is known to work with a  few percent accuracy.

In the case of the deep inelastic
scattering off the deuteron the only quantity which enters
into consideration is the simultaneous interaction of
the virtual photon with two nucleons. In this case
\begin{equation}
\sigma_{shad}^{D}
\equiv \sigma_{\gamma^*p} +\sigma_{\gamma^*n}- \sigma_{\gamma^*D},
\label{sigshad}
\end{equation}
is directly expressed through the cross section of diffraction in
the $\gamma^*N$ processes. (To simplify notations we do not write explicitly
in eq.\ref{sigshad} differential in $x$ and $Q^2$.) One has
\cite{Gribov}
\begin{equation}
\sigma_{shad}^{D}=\eta\frac{1}{4\pi}\int
 S(4t) \frac{d^2 \sigma^{\gamma^* + N \rightarrow X + N}}
{dM^2 dt} dM^2 dt \ .
\label{griboveq}
\end{equation}
Here $d^2 \sigma / dM^2 dt$ is the cross section for
producing a state
$X$ with mass $M$ and four-momentum transfer squared $t$ in the
inclusive diffractive
reaction $\gamma^* + N \rightarrow X +N $. $S(t)$ is the
electromagnetic form factor of the deuteron,
$-t=k_t^2+(xm_N(1+M_X^2/Q^2)+k_t^2/2m_N)^2$.
  For simplicity we neglect spin effects
in the wave function of the deuteron. The factor $\eta$
is due to nonzero value of the real part of the amplitude
$A$ of the diffractive processes,
\cite{FSAGK}:
$\eta={(1-\lambda^2)\over (1+\lambda^2)}$, and
\begin{equation}
\lambda =Re A/Im A \approx {\pi \over 2}{\partial
\ln (A/s) \over \partial \ln 1/x}.
\label{reim}
\end{equation}

The current data for cross section of
inclusive diffraction at HERA appear to indicate that
$\lambda=0.2-0.3$, leading to $\eta=0.8 \div 0.9$.
Note that in Eq.\ref{griboveq} there is no separation between
leading and nonleading twist effects. This shortcoming is
practically
unimportant to the extent that we restrict
ourselves by the leading twist effects only.

The account of the hard physics requires certain modifications of the Gribov
formulae for the limit of
small but
fixed $x$ and large $Q^2$.
The reason is that the hard QCD physics - emission of gluons leads to
the change of the relation between the coherence length and $x$ with
increase of $Q^2$.
For given $x$ average $l_{coh}$ decreases with increase of
$Q^2$ so that at extreemly large $Q^2$ contribution of
$l_{coh} \le 2 fm$ would dominate leading
to disappearance of nuclear shadowing. As a result in this limit
one becomes sensitive to the $A$-dependence of the nuclear parton
densities at $x\ge 0.02$: an enhancement of gluon and valence quark
parton densities at $x \sim 0.1$ and the EMC effect at $x\ge 0.4$.
As soon as the nuclear  parton densities at
$x \ge 0.02 $ are known at a low $Q^2$ normalization point this
effect can be taken into account using QCD DGLAP evolution
equations \cite{FS88,FLS}.
This leads to the QCD generalization of the Gribov formulae.
For heavier nuclei interactions with $N \ge 3$ nucleons
become important and hence one needs  a more detailed information about
diffraction to determine the shadowing effects.
Considerations of structure function of nuclei
$F_{2A}$ in the framework of the Gribov theory
were performed in a number of papers, see e.g.
\cite{FS88,FS89,FLS,BL,NZ,Piller,FSAGK,Kop,Barone,Orsay}
though effects of enhancement of gluon densities were neglected
in  \cite{BL,NZ,Piller,Kop,Barone,Orsay}.
Note in passing that the eikonal approximation maybe a
 good starting approximation for multiparton configurations. On the
otehr hand
if initial configuration consists of two bare
partons only inelastic interactions with two nucleons may contribute.

The phenomenon of nuclear shadowing for the quark sea
($F_{2A}(x,Q^2)$ as the leading twist effect
is well established experimentally, see e.g. \cite{data} though
in a rather limited range of DIS kinematics. Further studies of
this phenomenon which is one of the
key elements of the physics of $AA$ collisions at LHC
may be possible at HERA in a $eA$ mode \cite{report}.

It is possible to describe quantitatively these data by applying
the Gribov theory of nuclear shadowing\cite{Gribov} which connects
the shadowing for $F_{2A}(x,Q^2)$ and diffraction in DIS scattering
off a nucleon for the same $x,Q^2$
\cite{FS88,FS89,FLS,NZ,FSAGK,Kop,Orsay}. In the early papers
one had  to start with modeling the cross section of
diffraction in DIS which was done based essentially on the QCD
generalization of the Bjorken aligned jet model\cite{FS88}.
More recently one could employ information on
diffraction in the HERA kinematics to reduce the model
element of considerations \cite{FSAGK,Piller,Kop,Barone,Orsay}.
Overall uncertainties involved in the estimates of higher order
terms appear to be rather small - see discussion below.
At the same time the  effects leading to an enhancement of the parton
densities for $x \sim 0.02- 0.2$
which so far could be treated only on the basis of
the
momentum and baryon charge
sum rules \cite{FS88} prevents quantitative
predictions for $x \ge 0.01$.

It is worth noting emphasizing,
 that in the case of shadowing in $\gamma^*A$ reaction
one expects that there will be no smooth transition from  the 
real phton limit to the limit of small $x \le 10^{-3}$ and $Q^2\sim$
 few GeV$^2$ \cite{Poland} since the small $t$ 
diffraction for the real photon case in $2\div 3$ times larger that in the 
DIS limit. At the same time for larger $Q^2$ 
rapidity gap probability weakly depends on $Q^2$ leading to a weak dependence 
of shadowing on $Q^2$.

Note in passing that several approaches to nuclear shadowing
have been developed
based on the light-cone treatment of the nucleus wave function, for a recent
review and references see \cite{lcrev}. If one imposes within
such an approach a condition that the nucleus is build of weakly bound
nucleons, this  approach could be considered as
 complementary to the approach outlined above. However so far the
light-cone approaches discussed in\cite{lcrev}
did not deal with the diffractive physics and
hence it is not clear whether approximations currently
used  in these approaches are consistent with the  HERA diffractive data.

\section{Nuclear shadowing of charm quark and gluon distributions in nuclei}.

\subsection{Diffractive production of charm as a gluon partonometer}
Obviously if one would have the information on diffraction
in the scattering of the hard probe coupled directly to gluons -
"a gluon partonometer" - one would be able to repeat the same program for
the gluon shadowing.

The above mentioned observation of the
operational validity of the factorization
approximation for the range of $x,Q^2,M_X^2$ covered
by HERA \cite{ACW}
implies that we can actually select the processes
dominated by scattering off the gluon field of the target and use them for
calculating the gluon shadowing in nuclei.
We choose as such a process production of charm because  in
this case the QCD evolution equation
describes well the inclusive cross section while the factorization
approximation describes the diffractive charm production.
So we will apply the Gribov theory to calculate nuclear shadowing
for charm diffractive
production.  As far as the charm production of nuclei can be described
in terms of the QCD factorization theorem
we can interpret this shadowing as the shadowing of the nuclear gluon density.

To get a better insight into
the picture of nuclear shadowing it is convenient
to  consider it in the S-channel picture
in the nucleus rest frame
where the projectile is considered
as a superposition of  different quark-gluon  configurations. In
particular, in the case of the small $x$ charm production the
virtual photon can be represented as $c\bar c$, $c\bar c g$, $\ldots$
configurations which  scatter
off the target gluon field.
Since the $c\bar c$ configuration has a small size  $\le 1/m_c$,
diffraction in this picture occurs predominantly due to the
scattering of the colorless dipole
built of $c\bar c- g$ in a color octet configuration off the target
\cite{AFS,BuchHe}. In many respects these
configurations are analogous to the $q\bar q$ aligned jet configurations.
The cross section of the interaction of the color
octet dipole is larger by a factor of 9/4 than
that for the color triplet dipole
(for the same transverse separation). Hence
the relative importance of the diffraction for the
probes coupled to gluons and to quarks is  determined by the interplay of
two factors - larger cross section interaction for color octet dipoles
versus  the different probability of the
production of the large size configurations.

To characterize the dispersion over the strengths of
the interaction
it is convenient to introduce the notion of  cross section
fluctuations for the projectile nucleon
interactions. In our case we consider fluctuations of the
$c\bar c$ component of the
virtual photon wave function. The relevant quantity
is  the distribution over the
cross section strength $P(\sigma)$. The cross
section of diffraction at $t=0$ is expressed through $P(\sigma)$ using
the Miettenen-Pumplin relation\cite{MP}
generalized to account for the real part of the amplitude:
\begin{equation}
16 \pi{{d \sigma_{diff}\over dt}\over \sigma_{tot}}=
(1+\lambda^2){{<\sigma^2>} \over <\sigma> } \exp Bt,
\label{MPeqpr}
\end{equation}
where $\lambda=Re/Im $ was defined in eq.\ref{reim}.
By definition:
\begin{equation}
\left<\sigma^n\right>=\int d \sigma \sigma P(\sigma) \sigma^n
\end{equation}

The quantity which can be extracted
from the analysis of the data within
the model
of
\cite{ACW}
\begin{equation}
R_{diff}(charm)
={{\int_0^{x_0} dt dx_{\Pomeron}{d\sigma_{diff}^{charm}(M^2,s,x,Q^2)\over dt}}
\over \sigma_{tot}^{charm}(x,Q^2)},
\end{equation}
where $x_P=M^2/s$, $M^2$ is the mass$^2$ of diffractively
produced hadron system, and $ s$
is the square of the invariant energy of the $\gamma^*N$ collision.
We fix
$x_{o}=0.05$
since the nucleus form factor naturally
cuts the contribution of larger values of $x_{\Pomeron}$.
For the fit
{\bf D}
of \cite{ACW}
which provides the best description of
the charm data and dijet production in
 the real photon scattering
 we find $R_{diff}(charm)(x,Q^2)$
to be a very weak function of
$x,Q^2$ for $x\le 10^{-3}$ and $Q^2\ge 10 GeV^2$.
For the $10^{-4}\le x\le 2.5\cdot 10^{-3}$ range
the
ratio changes between from 0.20-0.21 at
$Q^2 \sim 8 GeV^2$ to
0.16-0.17 at $Q^2 \sim 70 GeV^2$.
Note that $R_{diff}(charm)$
 is larger than the corresponding
quantity for the total cross section
of diffraction in DIS where it is close to 0.10.
 Qualitatively this is
consistent with above mentioned larger cross
 section of interaction for the
color octet dipole. Recently the $t$
 dependence of the inclusive
diffractive cross section was measured
for $x_L \ge 0.97$ and $Q^2 \ge 3 GeV^2$
\cite{slopezeus}.
The data were fitted assuming
an
exponential dependence of the cross section on $t$:
${d \sigma_{diff}\over dt}
\propto \exp Bt$, finding $B= 7.1\pm 1\pm 1.2 GeV^{-2}$. This
 is consistent with
the dominance of strong interaction ` physics in
diffraction \footnote{Note that one
may expect that the $t$-dependence
of the diffractive cross section should
 be  better approximated for $-t\le 1 GeV^{-2}$ by
${d \sigma_{diff}\over dt}
\propto G_N^2(t) \exp {\tilde B}t$, where $G_N(t)$ is the
nucleon electric form factor.
Use of such a form would lead to
an increase of
${d \sigma_{diff}\over dt}_{\left|t=0\right.}$
and hence to a larger value of
$\sigma_{eff}$, cf.
Eq.\ref{sigmaeff}.}. Since the
diffraction forms a significant fraction of
total cross section of the charm production
the large size configurations should give large contribution
which is consistent with the
analysis of  \cite{ACW}
which assumes the Ingelman-Schlein model and the QCD evolution,
i.e assumes that
diffraction is due to a soft QCD-pomeron exchange.
Hence it is natural to assume that
the slope
$B$ of the diffractive production of charm
is the same as that
for the inclusive diffraction.
 Taking $R_{diff}=0.2$
and neglecting the real part of the amplitude
(we will explore sensitivity to the value of $\lambda$ below)
we find\footnote{In this
estimate we use the parameterization of \cite{ACW} for all
$\beta=Q^2/(Q^2+M^2)$, though for $\beta \sim 1$
higher twist effects are important. However this
region gives very small contribution to the total
diffractive cross section.}:
\begin{equation}
\sigma_{eff} \equiv \left<\sigma^2\right>/\left<\sigma\right>=
16\pi R_{diff}B=28 mb
\cdot
(1\pm 0.23)
\label{sigmaeff}
\end{equation}
To
illustrate the magnitude of the uncertainties we give  here
errors
due to uncertainties in the value  of the $t$-slope. We
do not include an error due to
uncertainties in the absolute cross section of diffraction
which is more difficult to estimate.
In the following analysis we will neglect a small decrease
of $R_{diff}$ with increase of $Q^2$ which leads
to a slow decrease of
the  shadowing with increase of $Q^2$.
We will consider this effect as well as the $x$ dependence of shadowing
 elsewhere \cite{FS982}.

The large value of $\sigma_{eff}$ for charm diffraction
is another indication that relatively soft 
QCD
physics determines the
cross section of the charm diffraction.
\footnote{$ \sigma_{eff}$ may be overestimated within the above evaluation
since one of characteristic features of  hard diffraction
is a smaller slope $B$. But a smaller slope according to
eq.\ref{sigmaeff} would lead to smaller $ \sigma_{eff}$.
 At the same time even for hard diffraction the slope is $\approx 4.5-5
GeV^{-2}$
as measured in the exclusive processes of $J/\psi$ and
$\rho$-meson production,
one would obtain a value of $\sigma_{eff} $ close to the lower limit of
eq.\ref{sigmaeff}. Moreover a better fit is given in this
 case by $f(t)=F_N^2(t)$, where $F_N(t)$ is the nucleon dipole form factor.
Using this fit one gets the values of $\sigma_{eff}$ within the range
of eq.\ref{sigmaeff}.}
It is worth emphasizing that there is no contradiction  between this
statement and the applicability of the QCD
factorization model. In this model this  statement
reflects the properties of the Pomeron parton density at
the $Q^2$ boundary
for the evolution. This is similar to the situation for $F_{2N}(x,Q^2)$.

\subsection{Quasieikonal approximation}
To evaluate
the shadowing for the scattering of an arbitrary nucleus we
have to sum the diagrams where the
projectile interacts with 2,3,4... nucleons.
The nuclear wave function enters via many-nucleon
 form factors, which
suppress large momentum transfer in the transition amplitudes.

First, let us make  a simplifying
assumption that all configurations
interact with the same
cross section $\sigma_{eff}$. This corresponds to
 neglecting
the dispersion of
the distribution over $\sigma$ in $P(\sigma)$ -
 we will explore this
assumption below. We also assume that the mass of
 the system
propagating between the points where
the photon has transformed into a
 hadron configuration and where the hadron
 configuration collapsed back
into a photon does not change by a large
 factor, say a factor
 larger than 2, or
 smaller than 0.5,
(see discussion of this point below).
In this case we can write an expression for
$R=\sigma(eA \to e +charm +X)/A
\sigma(eN \to e +charm +X)$
in the form analogous to the case of vector dominance model - eq.(5.4)
of \cite{VDM}. The only difference is the substitution
 $m_{\rho}^2/s \to (Q^2+M^2)/s \equiv x_{\Pomeron}$
 \cite{FS89} and
the integration over $x_{\Pomeron}$,
 i.e. over $M^2$-the mass of the
diffractively produced
system \footnote{ In the discussed approximation
$\left<\sigma^2\right>/\left<\sigma\right>$ is
 fixed and further effects of dispersion are
neglected. Hence it
is close to the enhanced eikonal approximation used in \cite{Kondr}}:
\ba
R & = & 1-{1 \over 2} Re \left[\sigma_{eff}(1-i\lambda)
\int d^2b\int_{-\infty}^{\infty}dz_1\int_{z_1}^{\infty} dz_2
\int_x^{x_0} dx_{\Pomeron}\eta
(x_{\Pomeron})
\rho_A(b,z_1)\rho_A(b,z_2)\right.
\nonumber \\
& &\left.\exp(-iq_{\parallel}(z_1-z_2))\exp(-\left({1\over 2}\sigma_{eff}
\int^{z_2}_{z_1}\rho_A(b,z')dz'\right)\right].
\label{master}
\ea

Here $\rho_A(r)$ is the nucleon density in the nucleus normalized according to
the equation $\int \rho_A(r)d^3r=1$. The quantity $\eta(x_{\Pomeron})$ is the
differential diffractive cross section
normalized to the total diffractive cross section
\begin{equation}
\eta(x_{\Pomeron})={{d\sigma_{diff}(x_{\Pomeron})\over dx_{\Pomeron}}
 \over \int_x^{x_0}{\sigma_{diff}(x_{\Pomeron})\over dx_{\Pomeron}}}
\end{equation}
 The parameter $q_{\parallel}=x_{\Pomeron}m_N$
is the longitudinal momentum transfer to the nucleon in the
 $\gamma^* \to M_X$ transition. It is simply related to
 the coherence length
parameter defined in eq.\ref{lcoh} as  $q_{\parallel}={1\over l_c}$.

The factor
$\exp(-iq_{\parallel}(z_1-z_2))$
in eq.\ref{master} suppresses the
contribution of large $x_{\Pomeron}$. To illustrate this point we
present in Fig.1 the result of the calculation of the ratio
$(1-R(x_{\Pomeron}))\over (1-R(x_{\Pomeron}=0))$
for the value of $\sigma_{eff}=28 mb$,
where instead of integrating
over $x_{\Pomeron}$ we fix its value
in eq.\ref{master}. One can
see from this plot that the suppression depends weakly on $x_{\Pomeron }$,
which makes the neglect of the
change of the mass quite safe. \footnote{This approximation is
similar to the
one used in hadron-nucleus scattering in the model \cite{Kondr}
which describes well available data on $\sigma_{tot}(hA)$
for the incident hadron energies
$E^h_{inc} \le 400 GeV$
 where inelastic shadowing effects
play a noticeable role.}
One can see also
that the contribution of
the region of $x_{\Pomeron} \ge 0.05 $ is strongly suppressed.

To explore the sensitivity of the result to the value
of $\sigma_{eff}$
we present in Fig.2 the value of $A_{eff}/A$ calculated
for different values of $\sigma_{eff}$, and $\lambda=0$
 in the ranges relevant for shadowing
in the gluon channel and
in inclusive $eA$ scattering.
 One can see that the estimated uncertainties in
$\sigma_{eff}$ of $\sim 23\%$ do not lead to a
significant change of
the shadowing
effect. We will explore details of the $Q^2$ and $x$
dependence of gluon shadowing elsewhere.

Note also that the cross section of the diffractive charm
production rapidly increases with increase of energy. Hence
 eq.\ref{reim} implies that the real part of the amplitude
 is not small. Based on the fits of \cite{ACW} and eq.\ref{reim}
we estimate $\lambda \sim 0.4$. Hence we need to explore the
 sensitivity of $A_{eff}/A$ to the value of  $\lambda$.
In Fig.3 we present the result of calculation for
$A_{eff}/A$ for A=12,240 for a range of the values of $\lambda$
and
for the fixed value of $\sigma_{eff}(1+\lambda^2)$
which is determined from the diffractive ratio of eq.\ref{MPeqpr}.
One can see from the figure that for a realistic range of
$\lambda$ (shaded area) the sensitivity to real part is
 rather small, especially for heavy nuclei.

\subsection{Shadowing for central impact parameters}
For the heavy ion applications it is interesting to consider shadowing
of DIS processes
for the central impact parameters.
It is natural to define the parton density per unit area and to consider the
suppression factor $R(b)$ for the value  of this
density as compared to that in the  uncorrelated nucleons placed
at the same impact parameter. The same equation \ref{master} is
valid for $R(b)$
with the integral over $d^2b$ removed and an extra factor
$\int dz \rho(b,z)\equiv T(b)$
in the denominator \cite{FLSbnl}.
\ba
R(b) & = &
1-{A \over 2T_A(b)} Re \left[\sigma_{eff}(1-i\lambda)
\int_{-\infty}^{\infty}dz_1\int_{z_1}^{\infty} dz_2
dx_{\Pomeron}
\int_x^{x_0}\eta(x_{\Pomeron}\rho_A(b,z_1)\rho_A(b,z_2)\right.
\nonumber \\
& &\left.\exp(-iq_{\parallel}(z_1-z_2))\exp(-\left({1\over 2}\sigma_{eff}
\int^{z_2}_{z_1}\rho_A(b,z')dz'\right)\right].
\label{masterb=0}
\ea

 For large $A$ and $b=0$ neglecting the edge
effects (that is taking the nuclear density constant and equal
to $\rho_0 \approx 0.16 fm^{-3}$ we obtain neglecting
the $q_{\parallel}$ effects but accounting for the fluctuations of
the interaction strengths as:
\begin{equation}
A_{eff}(b=0)/A={\int d\sigma P(\sigma)
(2-2 \exp(-\sigma L\rho_0/2)) \over \rho_0L\int d\sigma \sigma P(\sigma)},
,
\label{b0}
\end{equation}
where $L \approx 2R_A$. If we
neglect
the dispersion in $\sigma $ this would lead, for
 $\sigma_{eff}L\rho_0/2 \gg 1$,
to 
\begin{equation}
A_{eff}(b=0)/A \approx {2\over \rho_0L\sigma_{eff}} 
\end{equation}
For $A=208$ this 
results in
$A_{eff}(b=0)/A=0.32$ which is
significantly smaller than the average $A_{eff}/A$ for lead. 

\subsection{Fluctuations of the interaction strength}
Let us investigate the
sensitivity of this estimate to dispersion of the interaction strength.
Motivated by the general structure of $P(\sigma)$ for a photon,
where for small $\sigma$ - $P(\sigma)\propto {1 \over \sigma}$ \cite{FRS}
(this behaviour is essentially consequence of the dimentional counting for 
$P(\sigma)$ for a point-like object),
we take a model:
\begin{equation}
P(\sigma)=c{1\over \sigma} \theta(\sigma_0-\sigma)
\end{equation}
The parameter $\sigma_0=2\sigma_{eff}$ is  fixed 
based on eq.\ref{sigmaeff}.
In this case eq.\ref{b0} leads to 
\begin{equation}
A_{eff}(b=0)/A={\int^{\sigma_0}_0(1-
\exp(-\sigma \rho_0L/2))
/\sigma d\sigma \over \sigma_0\rho_0 L/4}
\label{fl}
\end{equation}
One can see from the analysis of 
eq.\ref{fl} that for large values of 
the nuclear thickness the
shadowing effect is smaller when the fluctuations are included
due to the contribution of small $\sigma$'s. However numerically even
 for the
heaviest nuclei the effect is an
increase of $A_{eff}/A$ by a factor of 1.3,
see Fig.4.
The increase 
occurs due to 
the
contribution
of configurations with $\sigma \le \sigma_{eff}$. For very large
$A$ this would lead to a dominance of 
the contribution
of configurations with small $\sigma$ - a kind of a filtering 
phenomenon. To estimate at what $A$ small $\sigma$ would dominate
we present in Fig.5 
the fraction of the cross section due
to configurations with $\sigma$ smaller than a given one.
One can see that for $\sigma_{eff}=28$mb one would need extremely 
large $A$ 
values
to reach a situation where say
 $\sigma \le 10 $mb would give 
the
dominant contribution. 
Due to the condition $l_{coh} \gg 2 R_A$ this would require extremely small 
$x$.

The significant values of shadowing 
derived for small $x$ and a wide range of
$Q^2$ have obvious implications for $AA$ collisions at LHC. For example,
for production of $J/\Psi$ and $\Upsilon $ at 
the central rapidities in Au-Au collisions we
predict a suppression of the order of 1/5 ($\sim A^{-.3})$
for the inclusive cross section
and $\sim 1/8 (\sim A^{-.4})$ for the central collisions solely due
to the gluon shadowing effects. A similar
suppression should occur
for the minijet production for $x_T(jet) \le 10^{-3}$. 

Further suppression occurs due to the final state of the produced onium
states ($\chi,J/\psi,...$).

\subsection{Structure of the final states}
Let us briefly discuss also the structure
 of the final states for 
the processes dominated by scattering off 
the gluons for $x \le 10^{-3}$. 
In the case of inclusive scattering we have
demonstrated in \cite{FSAGK} using 
the AGK technique \cite{AGK}
that 
larger shadowing effects larger
the rapidity gap fraction of events and an
increase of multiplicity fluctuations
for production of soft particles for the central rapidity interval.
Since in the  case  of the processes dominated by coupling to gluons
$\sigma_{eff}$ is a factor $\sim 2$ 
larger than for the inclusive case - 
application of the same technique indicates that for the 
case of charm production
(i) the fraction of charm produced in the rapidity gap  events for heavy nuclei
should be in the range of
40-50\%, (ii) the
fraction of diffractive events with $\beta=Q^2/
(Q^2+M^2_{diff}) \ge 0.4$ should increase due to 
the enhancement
of "elastic" scattering of the projectile configurations as compared
to the "triple Pomeron type " processes \cite{FSAGK},
 (iii) the distribution over
the number of particles produced for a fixed rapidity interval around
$ y_{c.m.} \sim 0$ should be much broader than in $ep$ collisions.
Also, since the  amplitude of hard diffractive process like 
$\gamma_L+A \to V(\rho,J/\psi,...)+A$ is proportional to 
$x^2G_A^2(x,Q^2)$
for $t=0$ - 
(the generalized color transparency \cite{gct}), 
the cross section of this process integrated over $t$
should be $\sim A$, since $xG_A(x,Q^2)/A \propto A^{-0.15}$.
Since $\sigma_{tot}(eA)/A \propto A^{-0.09}$ \cite{FSAGK} we expect
 $\sigma(\gamma_L+A \to V(\rho,J/\psi,...)+A)/\sigma_{tot}(eA) \propto 
A^{0.1}$.

In conclusion, we have demonstrated that the observation of 
the large diffractive 
cross section of charm production at HERA indicates that the shadowing
for gluon densities at $x\le 10^{-3}$ should be large and strongly 
affect the physics of nucleus-nucleus collisions at LHC. 
The observation of gluon shadowing of such magnitudes and
 associated effects in diffraction 
processes would be one of the green pastures for the $eA$ collider program at
HERA.

\acknowledgments{One of us (M.S.) would like to thank DESY
for the hospitality during the time this work was done. 
We thank J.C.Collins, H.Jung and J.Whitmore
for discussion of the charm diffractive production.
We are indebted to A.Levy for reading paper
and for the valuable comments.
This work is supported in part by the U.S.
 Department of Energy and BSF.}

\newpage

\vspace{1.0cm}
\begin{figure}
\centerline{
\epsfig{file=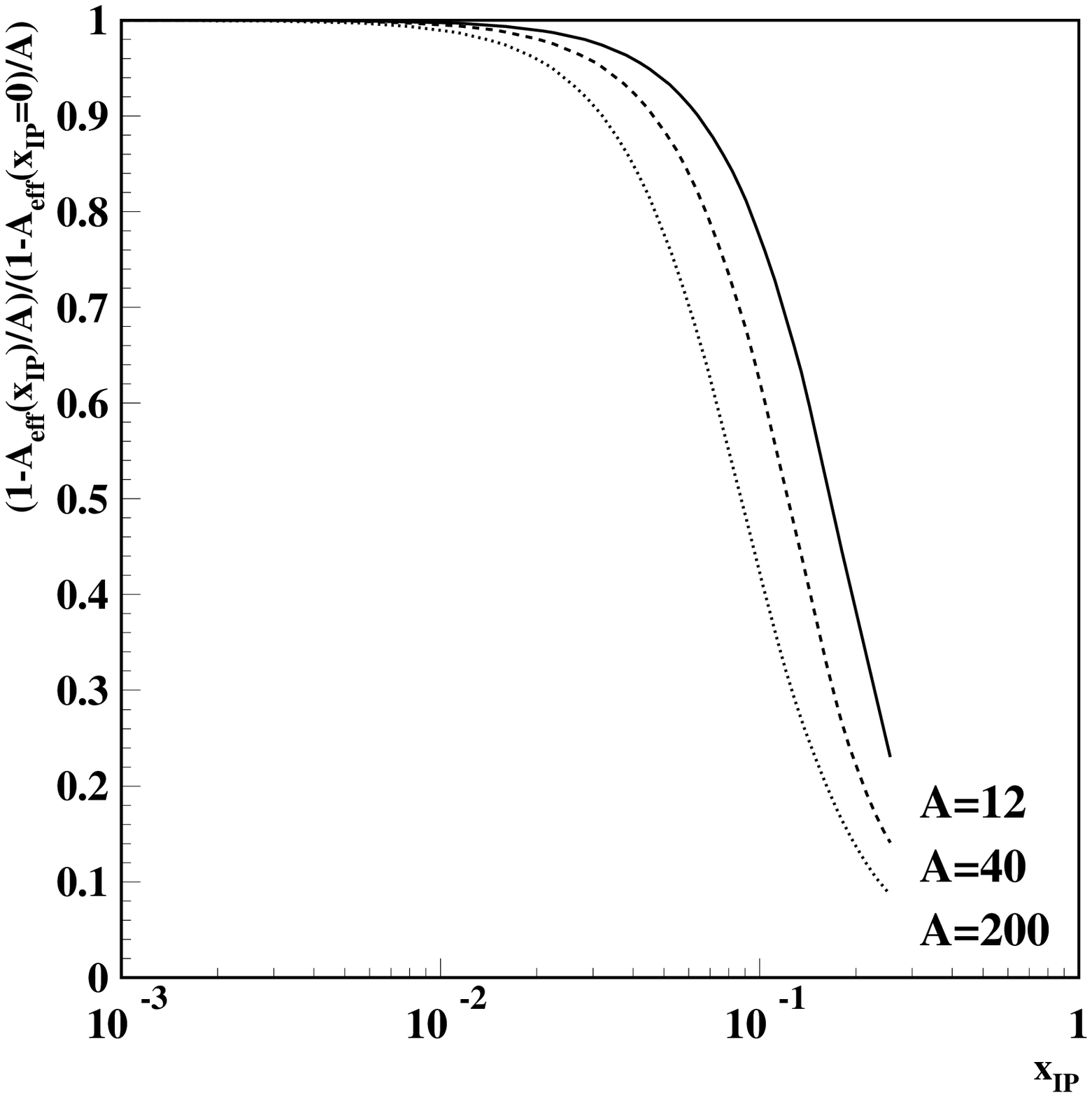,width=15.0cm,height=15.0cm}}
\caption{Suppression due to 
the
longitudinal momentum
 transfer calculated as a function of $x_{\Pomeron}$.}
\end{figure}

\newpage

\vspace{1.0cm}

\begin{figure}
\centerline{
\epsfig{file=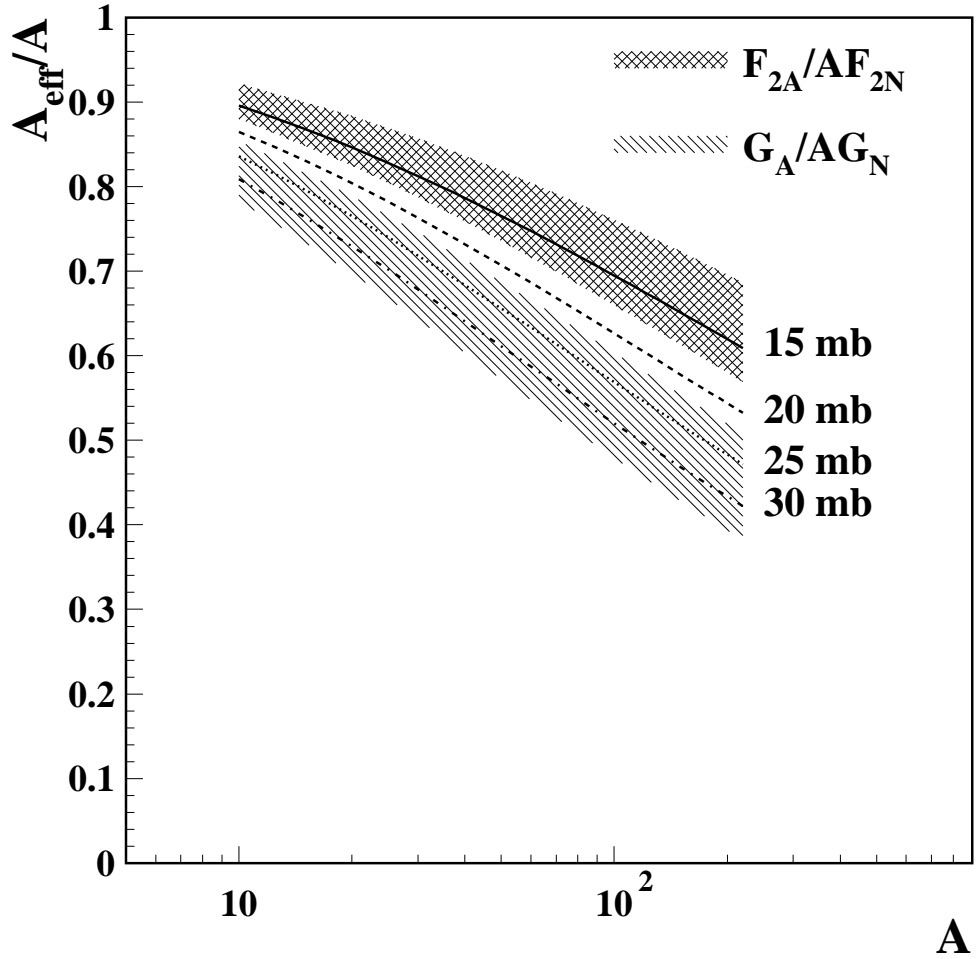,width=15.0cm,height=15.0cm}
}

\caption{Dependence of $A_{eff}/A$ on $\sigma_{eff}$ for
 $\lambda=0$. 
The 
shaded regions are 
expected values of $F_{2A}/AF_{2N}$ 
and $G_{A}/AG_{N}$ for the  range
of $\sigma_{eff}$
values
consistent with the HERA diffractive data.}
\end{figure}
\newpage

\vspace{1.0cm}

\begin{figure}
\centerline{
\epsfig{file=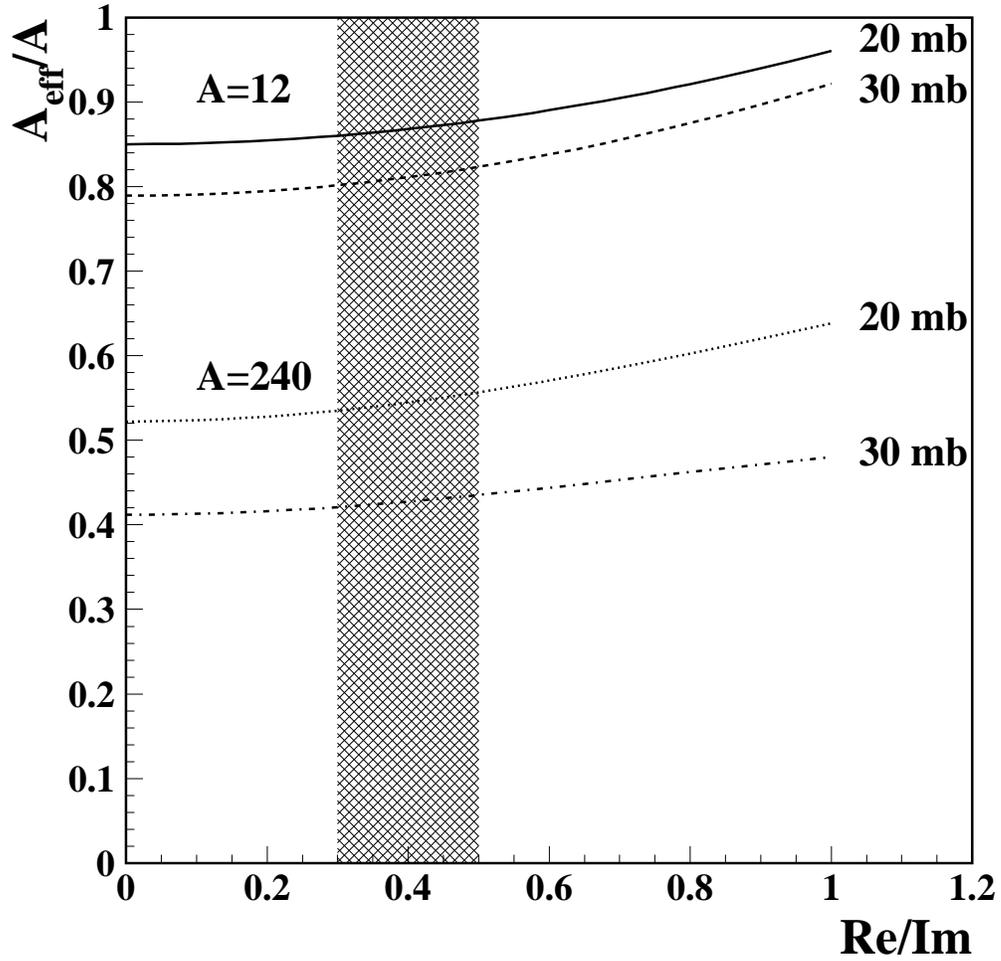,width=15.0cm,height=15.0cm}
}

\caption{Dependence of $A_{eff}/A$ on 
 $\lambda\equiv Re/Im$ for A=12 and A=240
and $\sigma_{eff}(1+\lambda^2)$ equal to 20 and 30mb. 
The 
shaded region indicates a realistic range of $\lambda$ for
charm diffraction.}
\end{figure}

\newpage

\vspace{1.0cm}

\begin{figure}
\centerline{
\epsfig{file=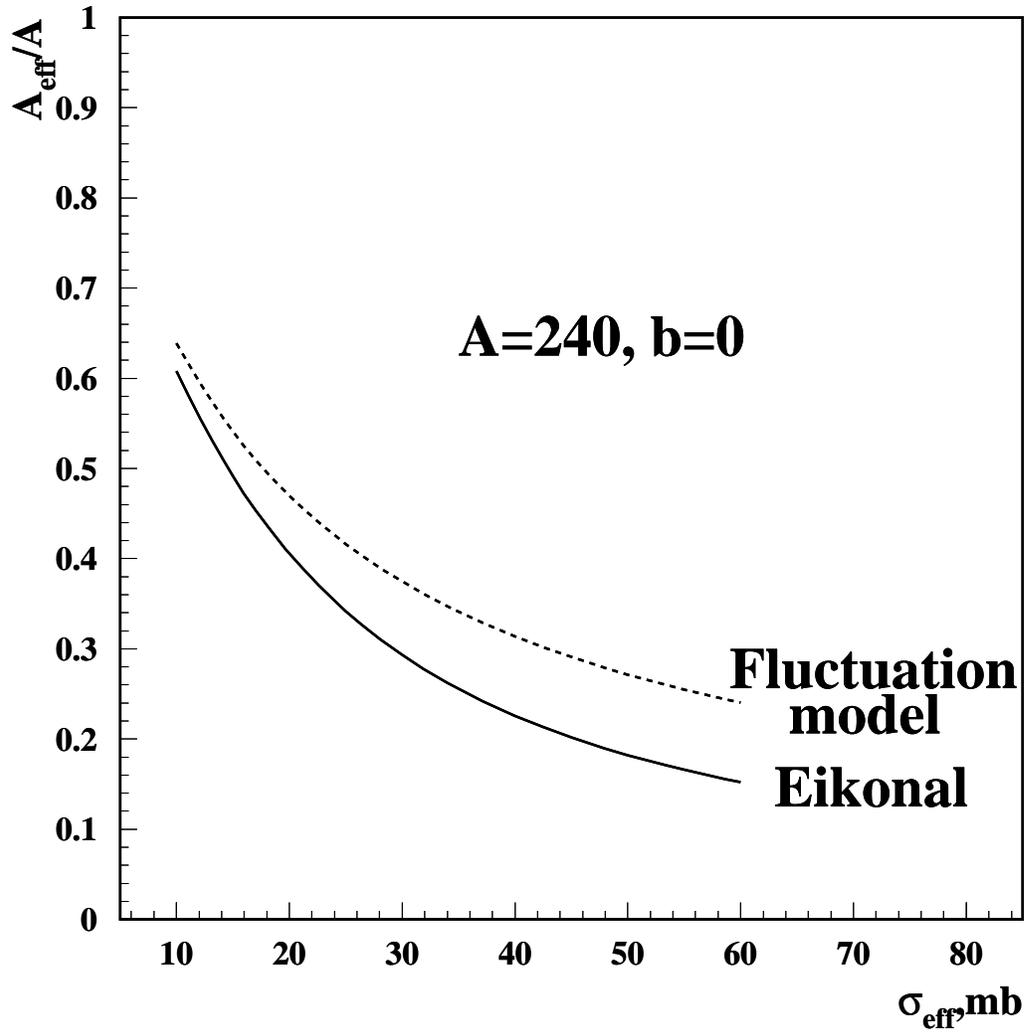,width=15cm,height=15cm}}

\vspace{1.0cm}

\caption{Comparison of the 
quasieikonal based model
and
the fluctuation model for the value of shadowing for A=240,
and $b=0$,
as a function of $\sigma_{eff}$.}
\end{figure}

\begin{figure}
\centerline{
\epsfig{file=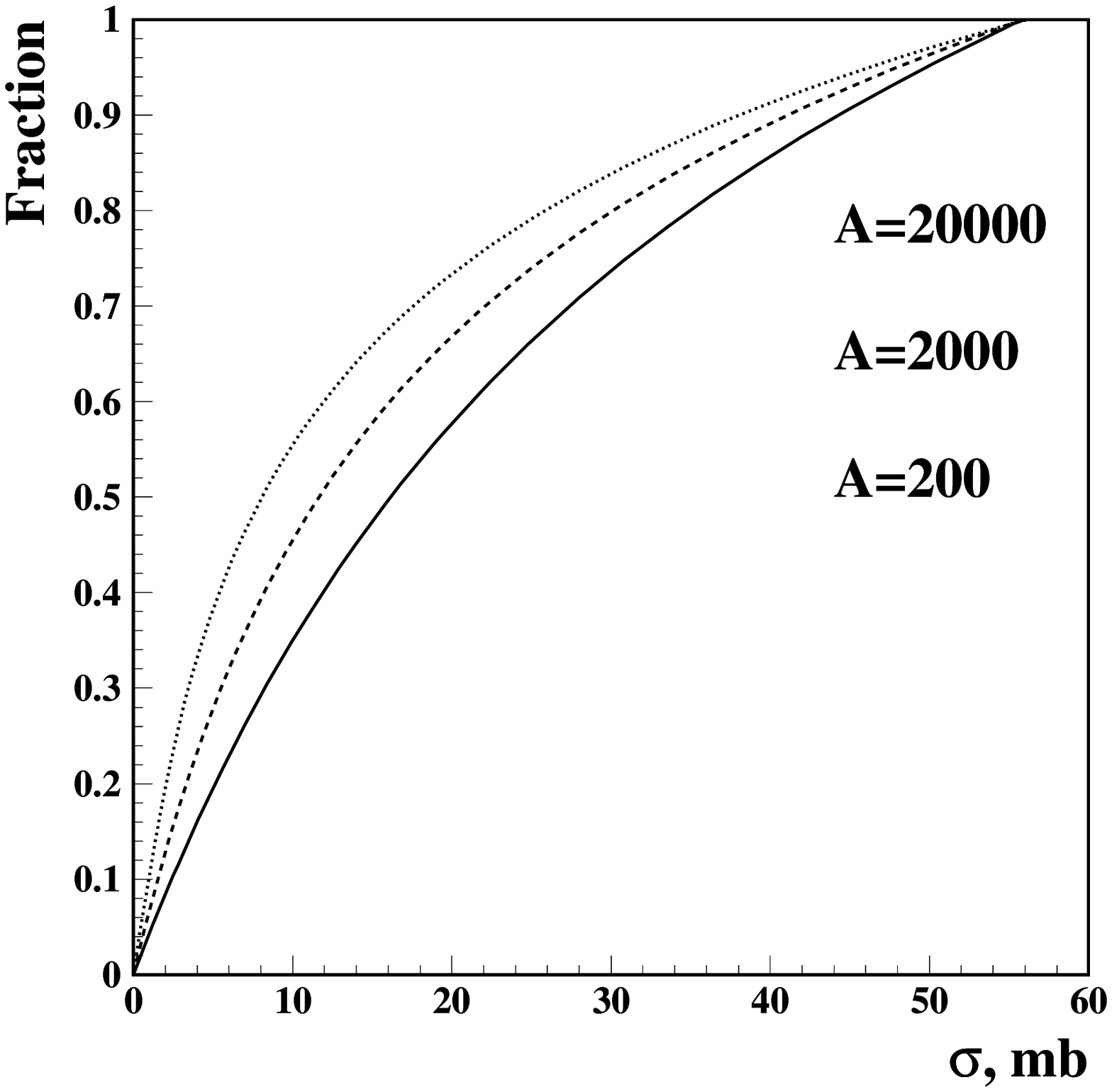,width=15cm,height=15cm}}
\caption{Fraction of the total cross section due
 to the contributions of 
cross sections $\le \sigma$ for $b=0$ and 
A=200(solid curve),A=2000(dashed curve) and A=20000(dotted curve)
.}
\end{figure}


\begin{thebibliography}{99}
\bibitem{Gribov} V.N.Gribov, Sov.J.Nucl.Phys. {\bf 9} 1969) 369; 
Sov.Phys.JETP {\bf 29} 1969, 483; ibid {\bf 30} (1970) 709;
\bibitem{HERAdiff}
  ZEUS Collaboration, J. Breitweg et al.,
    Eur.\ Phys.\ J. {\bf C1} (1998) 81,
    hep-ex/9709021.

    H1 Collaboration, C. Adloff et al.,
     Z. Phys.\ {\bf C76} (1997) 613,
     hep-ex/9708016.


  ZEUS Collaboration, ``Diffractive Dijet Cross Sections in 
    Photoproduction at HERA'', hep-ex/9804013.

H1 Collaboration, Diffractive Dijet Production at HERA,
 DESY-98-092,  submitted to Eur. Phys. J. C 

  H1 Collaboration, ``Production of $D^{*\pm}$ Mesons in
 Diffractive Interactions at HERA,  contribution 559 to XXIX
 International
 Conference on High Energy 
Physics, Vancouver, July 1998.

   ZEUS Collaboration, ``Open Charm Production
in Deep Inelastic diffractive scattering at HERA'',
    contribution 785 to XXIX International
 Conference on High Energy 
Physics, Vancouver, July 1998.
\bibitem{ACW}L. Alvero, J. C. Collins, J.J. Whitmore, 
hep-ph@xxx.lanl.gov - 9806340,  hep-ph@xxx.lanl.gov - 9805268.
\bibitem{FS982} L.Frankfurt and M.Strikman, hep-ph-9810331. 
\bibitem{ioffe}N. Gribov, B.L. Ioffe, I.Ya. Pomeranchuk,
 Sov.J.Nucl.Phys.6:427,1968,  Phys.Lett. {\bf 24B} (1967) 554;

B.L. Ioffe Phys.Lett. {\bf 30B} (1969) 123.

\bibitem{FS89}  L. L. Frankfurt and M.I.Strikman,
Nucl. Phys. {\bf B316} (1989)340. 
\bibitem{FSAGK} L.L. Frankfurt, M.I. Strikman,
Phys. Lett. 
{\bf B382} (1996) 6.
\bibitem{FS88} L. L. Frankfurt and M. Strikman, Phys.Rep. {\bf 160} (1988) 235.
\bibitem{FLS} L. L. Frankfurt, M. Strikman and S. Liuti, Phys. Rev.
 Lett. {\bf 65} (1990) 1725.
\bibitem{BL} S.J.Brodsky and H.J.Lu, Phys. Rev. Lett. {\bf 64} (1990) 1342.
\bibitem{NZ} N. N. Nikolaev and B. G. Zakharov, Z. Phys. {\bf C49} (1991) 607.

\bibitem{Piller}G. Piller, G. Niesler, W. Weise, Z. Phys. {\bf A358}
 (1997) 407.
\bibitem{Kop}B. Kopeliovich, B.Povh,
 Phys.Lett.
{\bf  B367}(1996) 329. 
\bibitem{Barone}V. Barone, M. Genovese,
Phys. Lett.{\bf  B412} 
(1997) 143.

\bibitem{Orsay}A. Capella, A. Kaidalov, C. Merino, 
D. Pertermann, J. Tran Thanh Van, Eur. Phys. J.{\bf C5}  (1998) 111.



\bibitem{data}M. Arneodo et al.,
 Nucl.Phys. {\bf B481}
(1996) 3;


 M.Arneodo, Physics Reports {\bf 240} (1994) 301.


\bibitem{report}
 M. Arneodo, A. Bialas, M.W. Krasny,, T. Sloan, M. Strikman,
In the proceedings of Workshop on Future Physics at HERA,
 Hamburg, 887
(1996). 
\bibitem{Poland} M.Strikman, Acta Pysica Polonica, 
\bibitem{lcrev}R. Venugopalan nucl-th- 9808023, 1998.
\bibitem{AFS} H. Abramowicz, L.L. Frankfurt, and M. Strikman DESY-95-047, 
 March 1995; Proceedings of SLAC Summer Inst., 1994, pp. 539-574.  
\bibitem{BuchHe}W. Buchmuller, M.F. McDermott, A. Hebecker, 
Phys.Lett. {\bf B404} (1997) 353.
\bibitem{MP} H. Miettinen and J. Pumplin, Phys. Rev. {\bf D18} 
(1978) 1696.
\bibitem{slopezeus}ZEUS collaboration, Contributed paper 972,
 XXIX International Conference on HIgh Energy Physics,
 Vancouver, 23-29 July 1998.
\bibitem{VDM}T.H. Bauer, R.D. Spital, D.R. Yennie, F.M. Pipkin,
 Rev.Mod.Phys.{\bf 50}
(1978) 261, ERRATUM-ibid. {\bf 51} (1979) 407.
\bibitem{Kondr}V.A.Karmanov and L.A.Kondratyuk, JETP letters, {\bf 18}
 (1973) 266.
\bibitem{FLSbnl}L.L. Frankfurt, M.I. Strikman and S. Liuti,
 in Proceedings of 4 BNL Workshop on Relativistic Heavy Ion Collisions, 
July 1990, BNL 52262, p. 103-118.
\bibitem{FRS}L. Frankfurt, A. Radyushkin, M. Strikman,
 Phys.Rev.{\bf D55} (1997) 98;

L. Frankfurt, V. Guzey, and M. Strikman, Phys.Rev. {\bf D58} (1998) 094039.
\bibitem{AGK} V.Abramovski$\breve{{\rm i}}$, V. N. Gribov, and O. V.
 Kancheli, Sov. J. Nucl. Phys. {\bf 18}, (1974) 308. 
\bibitem{gct}L. Frankfurt, G.A. Miller and M. Strikman, Phys. Lett.
 {\bf B304}~(1993)~1;
S. J. Brodsky, L. Frankfurt, J.F. Gunion,
 A.H. Mueller, M. Strikman, Phys.Rev. {\bf D50}
 (1994) 3134-3144.


\end{thebibliography}
\end{document}